\documentclass[twocolumn,tighten]{aastex63}
\usepackage{graphicx}
\usepackage{amsmath}
\usepackage{scalerel}
\usepackage{apjfonts}
\usepackage{tikz}
\usetikzlibrary{svg.path}

\definecolor{orcidlogocol}{HTML}{A6CE39}
\tikzset{
  orcidlogo/.pic={
    \fill[orcidlogocol] svg{M256,128c0,70.7-57.3,128-128,128C57.3,256,0,198.7,0,128C0,57.3,57.3,0,128,0C198.7,0,256,57.3,256,128z};
    \fill[white] svg{M86.3,186.2H70.9V79.1h15.4v48.4V186.2z}
                 svg{M108.9,79.1h41.6c39.6,0,57,28.3,57,53.6c0,27.5-21.5,53.6-56.8,53.6h-41.8V79.1z M124.3,172.4h24.5c34.9,0,42.9-26.5,42.9-39.7c0-21.5-13.7-39.7-43.7-39.7h-23.7V172.4z}
                 svg{M88.7,56.8c0,5.5-4.5,10.1-10.1,10.1c-5.6,0-10.1-4.6-10.1-10.1c0-5.6,4.5-10.1,10.1-10.1C84.2,46.7,88.7,51.3,88.7,56.8z};
  }
}

\newcommand\orcidicon[1]{\href{https://orcid.org/#1}{\mbox{\scalerel*{
\begin{tikzpicture}[yscale=-1,transform shape]
\pic{orcidlogo};
\end{tikzpicture}
}{|}}}}

\usepackage{comment}
\usepackage{array,multirow,color,tablefootnote}
\usepackage{xcolor}

\begin{document}

\title{Photometric Confirmation and Characterization of the Ennomos Collisional Family in the Jupiter Trojans}


\author[0000-0001-9665-8429]{Ian~Wong}
\altaffiliation{NASA Postdoctoral Program Fellow}
\affiliation{NASA Goddard Space Flight Center, 8800 Greenbelt Road, Greenbelt, MD 20771, USA}
\correspondingauthor{Ian Wong}
\email{ian.wong@nasa.gov}

\author[0000-0002-8255-0545]{Michael~E.~Brown}
\affiliation{Division of Geological and Planetary Sciences, California Institute of Technology, Pasadena, CA 91125, USA}


\begin{abstract}
Collisional families offer a unique window into the interior composition of asteroid populations. Previous dynamical studies of the Jupiter Trojans have uncovered a handful of potential collisional families, two of which have been subsequently confirmed through spectral characterization. In this paper, we present new multiband photometric observations of the proposed Ennomos family and derive precise $g-i$ colors of 75 candidate family members. While the majority of the targets have visible colors that are indistinguishable from background objects, we identify 13 objects with closely grouped dynamical properties that have significantly bluer colors. We determine that the true Ennomos collisional family is tightly confined to $a'_{p} > 5.29$ au and $0.45 < \sin{i_{p}} < 0.47$, and the majority of its confirmed members have near-solar spectral slopes, including some of the bluest objects hitherto discovered in the Trojan population. The property of distinctly neutral colors that is shared by both the Ennomos family and the previously characterized Eurybates family indicates that the spectral properties of freshly exposed surfaces in the Jupiter region are markedly different than the surfaces of uncollided Trojans. This implies that the processes of ice sublimation and space weathering at 5.2 au yield a distinct regolith chemistry from the primordial environment within which the Trojans were initially accreted. It also suggests that the Trojans were emplaced in their present-day location from elsewhere sometime after the initial population formed, which is a key prediction of recent dynamical instability models of solar system evolution.
\end{abstract}

\section{Introduction}
\label{sec:intro}

The origin of the Jupiter Trojans has remained an enduring mystery for planetary science. These objects, which lie in the 1:1 mean motion resonance with Jupiter and are confined to two swarms centered on the L4 and L5 Lagrangian points, may hold the key to answering fundamental questions about the Solar System's formation and evolution. In the past few decades, a new paradigm of early solar system history has emerged that describes a period of rapid dynamical restructuring following the end of planet formation, which significantly altered the orbital architecture of the giant planets \citep[e.g.,][]{gomes2005,morbidelli2005,tsiganis2005,nesvorny2012,nesvorny2018}. While these theories were primarily developed to explain other observed phenomena, such as the unexpectedly high orbital eccentricities of the gas giants and the dynamically excited state of the Kuiper belt \citep[e.g.,][]{tsiganis2005,levison2008,levison2011}, most of them encapsulate the same essential corollary in relation to the Trojans --- namely, that these objects initially formed beyond the primordial orbits of the ice giants, and were subsequently scattered inward and captured by Jupiter during the dynamical instability \citep{morbidelli2005,nesvorny2013,roig2015}. The idea of a common origin for the Jupiter-resonant asteroid populations and the Kuiper belt has motivated efforts to characterize the surface composition of the Trojans. However, unambiguous spectroscopic evidence for such a scenario has hitherto been elusive \citep[e.g.,][]{dotto2006,emery2006,fornasier2007,melita2008,emery2011,yang2011,brown2016,sharkey2019}.

While these observational endeavors have yet to achieve a definitive resolution of the Trojans' evolutionary history, they have led to a consequential realization regarding the ensemble properties of these small bodies. Visible photometry of the Trojans that has been provided by the Sloan Digital Sky Survey (SDSS) has revealed a marked bimodality in the measured surface colors \citep{szabo2007,roig2008,wong2014}. Spectroscopy and photometry in the near-ultraviolet \citep{wong2019}, visible \citep[e.g.,][]{roig2008}, and near-infrared \citep[e.g.,][]{emery2011,grav2012} have corroborated this finding, demonstrating that the Trojans can be separated into two broad classes that differ systematically in their surface properties --- the so-called less-red and red subpopulations. This discovery has opened up a new dimension in our efforts to understand the Trojan population. With the revelation of an analogous color bimodality among Kuiper belt objects \citep[e.g.,][]{fraser2012,lacerda2014,wong2017,pike2017,fraser2022}, several theories have been proposed to explain the emergence of such inhomogeneities \citep[e.g.,][]{wong2016,nesvorny2020}. The burgeoning interest in Trojans has culminated in the approval and launch of NASA's Lucy flyby mission \citep{levison2021}, which will carry out intensive in situ observations of five Trojan systems between 2027 and 2033.

Collisional families provide a novel perspective on the formation and evolution of Trojans. Due to their small size (i.e., diameters less than 250 km), Trojans are expected to be undifferentiated bodies with a homogeneous bulk composition. It follows that the fragments that are produced by a catastrophic impact should share similar spectral properties \citep[e.g.,][]{cellino2002,masiero2015}. Crucially, collisional families provide a glimpse into the interior composition of these objects, as well as the effects of space weathering at their current location. A comparison of the collisional fragments with the background population may reveal systematic differences, which would in turn indicate that the present-day environment at 5.2 au is significantly different than the conditions within which the Trojans initially formed \citep{wong2016}. This finding would provide strong evidence for the dynamical instability hypothesis of solar system evolution.

Tthousands of new Trojans have been discovered over the past few decades. Consequently, repeated numerical investigations have searched for clusters of dynamically similar objects that may have been produced from singular impact events \citep[e.g.,][]{milani1993,beauge2001,roig2008,broz2011,nesvorny2015,vinogradova2015,rozehnal2016,holt2020b}. The most recent works in this area list six candidate collisional families \citep{nesvorny2015,rozehnal2016}: four in the L4 cloud, and two in the L5 cloud. Follow-up observations of two of these families have confirmed the surface color homogeneity of the constituent objects, thereby validating them as collisional families. The Hektor family in the L4 cloud are associated with the largest Trojan ($D \sim 250$ km) and predominantly consist of small fragments, which indicates that the generating impact was a cratering event as opposed to a shattering collision \citep{rozehnal2016}. Photometry of the family members has revealed uniform colors that are consistent with the red subpopulation \citep{roig2008,rozehnal2016}. However, given the cratering nature of the impact, it is expected that the fragments are composed primarily of material from the outermost layer of the progenitor body, which may be dominated by weathered regolith (e.g., an irradiation mantle; \citealt{wong2016}). Therefore, this family may not provide a diagnostic window into the bulk properties of the pristine interior.

The Eurybates family was the first confirmed family that arose from a shattering impact. This family is situated at relatively low inclinations within the L4 swarm and are the most numerous of the identified dynamical families. Several observing programs have obtained photometry and spectroscopy of the candidate family members, revealing a unimodal color distribution \citep{fornasier2007,roig2008,deluise2010,broz2011}. Notably, the average color of the confirmed Eurybates family members is somewhat bluer than either the less-red or the red background subpopulation. This finding has broad implications for the evolution of Trojans. The discovery of distinct surface colors suggests that the collisional fragments experienced a different thermal processing and space weathering history than the primordial, uncollided Trojans. This offers tentative support to the hypothesis that the present-day Trojan population was implanted from another region of the solar system. However, given that the Eurybates family is, to date, the only characterized collection of collisional fragments (formed from a catastrophic impact), the confirmation of additional candidate families is required before a broader trend of photometric distinctness among Trojan collisional families can be established.

To this end, we have carried out a major observational campaign to validate and characterize another dynamical family: the Ennomos family. Unlike the Hektor and Eurybates families, this family is located at relatively high inclinations within the L5 swarm \citep[e.g.,][]{broz2011,nesvorny2015,rozehnal2016}. In this paper, we present the findings of our work. Section~\ref{sec:targets} describes the various family member catalogs in the literature and the target list for our campaign. The details of our observations and photometric extraction are described in Section~\ref{sec:obs}. In Section~\ref{sec:ana}, we utilize the measured color distribution to determine the list of confirmed Ennomos family members, analyze their magnitude distribution in relation to the background population and the Eurybates family, and discuss possible fruitful avenues for future study. Finally, we briefly recap the main findings of the paper in Section~\ref{sec:con}.

\section{The Proposed Ennomos Family}
\label{sec:targets}

\begin{figure*}[t]
\begin{center}
\includegraphics[width=\linewidth]{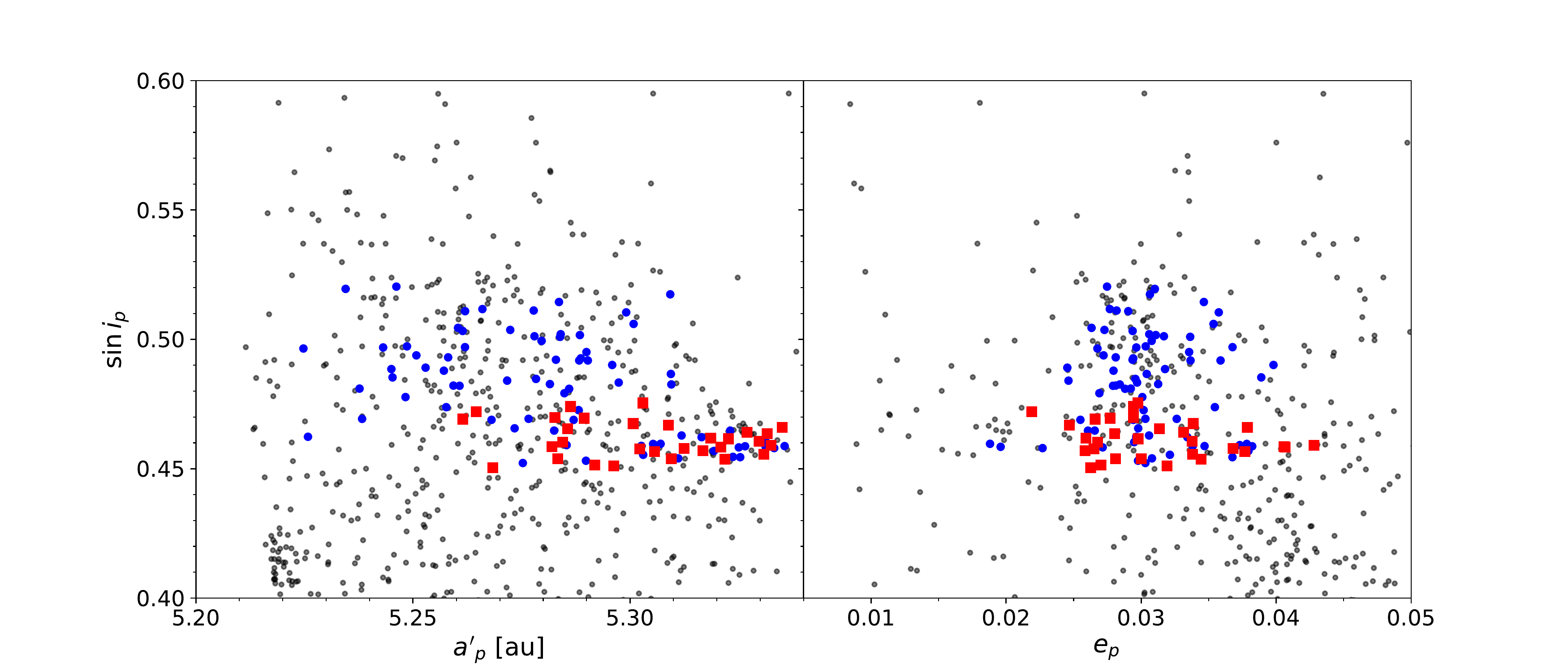}
\end{center}
\caption{A portion of the distribution of L5 Trojan proper elements $(a^{\prime}_{p},e_{p},\sin{i}_{p}$) taken from \citet{holt2020b}, where $a^{\prime}_{p}$ is the modified semimajor axis, defined in Section~\ref{sec:targets}. Background objects are shown in gray. The blue dots and red squares correspond to the Ennomos family members that are tabulated by \citet{nesvorny2015} and \citet{rozehnal2016}, respectively. The latter grouping, which served as the master target list for our photometric observations, contains significantly more objects and extends to higher proper inclinations and lower proper semimajor axes.} \label{fig:targets}
\end{figure*}

Asteroid collisional families are typically identified using numerical clustering algorithms, which define a distance metric between objects in the population and then search for statistically significant overdensities in the space of the input parameters; see, for example, the review by \citet{nesvorny2015}. The most common parameters that are used in clustering analyses are the proper orbital elements --- semimajor axis $a_p$, eccentricity $e_p$, and inclination $i_p$ \citep[e.g.,][]{milani1992,milani1994,knezevic2002}. Unlike the instantaneous osculating orbital elements, which vary periodically due to the perturbing gravitational effects of the giant planets, the proper elements have been corrected for these temporal variations and therefore encode the long-term dynamical state of each object. Because the ejection velocities of collisional fragments are usually much smaller than the orbital speed of the parent body, collisional families are expected to remain clustered in proper element space across timescales that are comparable to the age of the solar system.

The unique three-body dynamics of Trojans impose special conditions on the treatment of proper elements. The Trojans are confined to two swarms that are centered on the L4 and L5 Lagrangian points, with orbits that librate periodically. As such, the modified semimajor axis $a^{\prime}_{p} = a_{\rm cent} + D$ is often used, where $a_{\rm cent}$ is the semimajor axis of the center of libration, and $D$ is the libration amplitude $D = {\rm max}[a(t) - a_{\rm Jup}(t)]$ \citep[e.g.,][]{milani1993,broz2011,rozehnal2016}; here, $a$ and $a_{\rm Jup}$ are the osculating semimajor axes of the Trojan and Jupiter, respectively. The standard metric $d$ of \citet{zappala1994} is then used to quantify the distance between pairs of Trojans in the space of proper elements. Once a candidate collisional family is identified by the clustering algorithm, the effective spatial extent is set by the cutoff distance $d_{\rm cut}$, which is in units of m~s$^{-1}$. 


Several Trojan collisional family catalogs have previously been published, with each one utilizing a different orbital element database, cutoff distance, and/or cluster identification methodology \citep[e.g.,][]{milani1993,beauge2001,broz2011,nesvorny2015,vinogradova2015,rozehnal2016}.  The earliest published list of Trojan families that includes the proposed Ennomos collisional family was compiled by \citet{nesvorny2015} and is maintained on the Planetary Data System Small Bodies Node.\footnote{\url{https://sbn.psi.edu/pds/resource/nesvornyfam.html}} Using the custom Trojan proper elements that were computed by \citet{broz2011} as input, the Hierarchical Clustering Method \citep{zappala1994} identified a tight grouping of 30 objects in the L5 swarm with $d_{\rm cut} = 100$ m~s$^{-1}$, confined to $a^{\prime}_{p}\in\left\lbrack5.25,5.34\right\rbrack$ au, $\sin{i}_{p}\in\left\lbrack0.45,0.48\right\rbrack$, and $e_{p}\in\left\lbrack0.020,0.045\right\rbrack$. More recently, \citet{rozehnal2016} calculated updated proper elements following the methods of \citet{broz2011} for an expanded set of Trojans and applied a novel `randombox' Monte Carlo cluster identification technique. Their analysis yielded a significantly larger list of candidate Ennomos family members with $d_{\rm cut} = 100$ m~s$^{-1}$ consisting of 104 objects that extend to higher proper inclinations and smaller proper semimajor axes.\footnote{\url{https://sirrah.troja.mff.cuni.cz/~mira/mp/fams/}} Many of the additional candidate objects are fainter and were not included in the proper element database used in \citet{nesvorny2015}. Notably, this newer tabulation designates 17492 Hippasos as the center of the collisional family, instead of 4709 Ennomos, while 1867 Deiphobus is the largest member. 

Figure~\ref{fig:targets} shows a portion of the proper element distribution for L5 Trojans. The proper element values in this plot and hereafter are taken from \citet{holt2020b}, which presented the newest catalog that is currently available in the literature. The Ennomos family members that are listed in the works cited above are marked by the colored symbols. When planning our observations, we took the larger set of candidate objects from \citet{rozehnal2016} as the operative target selection list.

\section{Observations}
\label{sec:obs}
To measure the colors of potential Ennomos family members, we carried out multiband photometry of dozens of targets during two observing runs in 2015/2016 and 2021. A summary of the observations is provided in the following subsections. We also include a brief overview of the data processing and photometric extraction methodology.

\subsection{Keck/DEIMOS}
\label{subsec:keck}

\begin{deluxetable}{lcccc}[t!]
\setlength{\tabcolsep}{6pt}
\tablewidth{0pc}
\renewcommand{\arraystretch}{0.9}
\tabletypesize{\footnotesize}
\tablecaption{
    Summary of Keck/DEIMOS Observations
    \label{tab:deimos}
}
\tablehead{
    \colhead{Object} & \colhead{UT Date} & \colhead{UT Time\tablenotemark{\scriptsize a}} & \colhead{Airmass\tablenotemark{\scriptsize a}} &
    \colhead{$D_{\mathrm{aper}}$\tablenotemark{\scriptsize b}}   
}
\startdata
4709 & 2015 Dec 6 & 13:08:35 & 1.06 & 23 \\
17492 & 2015 Dec 6 & 13:46:40 & 1.04 & 23, 23 \\
32461 & 2015 Dec 6 & 10:47:25 & 1.38 & 17, 19 \\
36624 & 2015 Dec 6 & 15:02:17 & 1.63 & 17 \\
36624 & 2016 Jan 9 & 09:13:09 & 1.13 & 15 \\
48373 & 2015 Dec 6 & 10:53:33 & 1.73 & 15 \\
55419 & 2015 Dec 6 & 11:33:42 & 1.03 & 21, 25 \\
69437 & 2015 Dec 6 & 13:40:24 & 1.01 & 13, 23 \\
76867 & 2016 Jan 9 & 09:33:30 & 1.19 &  13 \\
77894 & 2015 Dec 6 & 12:25:36 & 1.00 & 17, 19 \\
98362 & 2015 Dec 6 & 13:26:59 & 1.00 & 13, 17 \\
122592 & 2015 Dec 6 & 10:41:33 & 1.32 & 17, 19 \\
154417 & 2015 Dec 6 & 12:04:47 & 1.01 & 19, 19 \\
187692 & 2016 Jan 9 & 08:31:05 & 1.24 & 13 \\
188976 & 2015 Dec 6 & 12:16:52 & 1.15 & 23, 25 \\
215319 & 2015 Dec 6 & 11:39:33 & 1.03 & 19, 23 \\
215542 & 2015 Dec 6 & 15:28:48 & 1.46 & 21, 17 \\
246817 & 2015 Dec 6 & 11:26:30 & 1.13 & 17, 21 \\
247967 & 2015 Dec 6 & 14:48:19 & 1.03 & 25 \\
284226 & 2015 Dec 6 & 15:22:51 & 1.32 & 17, 19 \\
287454 & 2015 Dec 6 & 13:33:16 & 1.02 & 15, 17 \\
293486 & 2016 Jan 9 & 09:01:30 & 1.22 & 15 \\
321706 & 2015 Dec 6 & 15:08:43 & 1.61 & 15 \\
337420 & 2016 Jan 9 & 08:51:34 & 1.24 & 15 \\
2006 BK240 & 2016 Jan 9 & 09:43:31 & 1.21 & 13 \\
2008 FD133 & 2016 Jan 9 & 10:03:32 & 1.17 & 13 \\
2012 QK22 & 2016 Jan 9 & 09:53:42 & 1.25 & 15 \\
\enddata
\vspace{+0cm}\textbf{Notes.}
\vspace{-0.25cm}\tablenotetext{\textrm{a}}{For each object, the start time and airmass value of the first exposure of the observing sequence(s) are provided. A 60 s exposure time was used for all images.}
\vspace{-0.25cm}\tablenotetext{\textrm{b}}{The diameter of the circular photometric aperture (in pixels) that yields the smallest resultant color uncertainty. The pixel scale of the unbinned DEIMOS images is $0\overset{''}{.}1185$. For objects that were observed twice on 2015 December 6, the optimal aperture diameters for the two visits are listed separately (see the text for details).}
\vspace{-0.8cm}
\end{deluxetable}

We observed 27 Trojans on UT 2015 December 6 and 2016 January 9 using the Deep Imaging Multi-object Spectrograph (DEIMOS) instrument on the Keck II telescope at Maunakea, Hawaii. DEIMOS is an optical spectrograph that is optimized for faint-object observations, with a detector mosaic consisting of eight $2048 \times 4096$ CCDs. In direct imaging mode, only four of the CCDs are used, with a reduced field of view of $16\overset{^{\prime}}{.}7 \times 5\overset{^{\prime}}{.}0$ and a pixel scale of $0\overset{''}{.}1185$. Each CCD features two amplifiers on opposite sides, and readout proceeds in parallel, resulting in two image extensions per CCD (i.e., a total of eight stored data arrays in each FITS file). We selected the standard telescope pointing for imaging observations, which placed the target in the middle of CCD \#3 near the boundary between the fifth and sixth image readout regions.

DEIMOS is equipped with the standard set of $BVRIZ$ filters. We imaged each field with the $V$ (5452 \AA), $R$ (6487 \AA), and $I$ (8390 \AA) filters while using sidereal tracking. A uniform per-exposure integration time of 60 s was applied to all science observations, which ensured a sufficient signal-to-noise ratio in all bands and negligible target motion during each exposure. For the night of 2015 December 6, the filters were cycled in the order $V$--$R$--$I$ at each pointing, and most targets were observed twice during the night. The targets that were visited on 2016 January 9 were only observed once using the filter sequence $V$--$R$--$I$--$I$--$R$--$V$. A series of bias frames and dome flats were collected at the start of each night. The observing conditions were excellent on both nights, with little to no cloud cover and measured seeing values in the range $0\overset{''}{.}6$--$1\overset{''}{.}0$.

Table~\ref{tab:deimos} lists the Trojans that we observed with Keck/DEIMOS, along with the start time and target airmass of the first exposure for each object and the photometric extraction aperture that was used for each visit (see Section~\ref{subsec:phot} for details). The only object that was observed on both nights was 36624 (2000 QA157). A handful of observations suffered from blending between the Trojan target and background stars, or were saturated, including all exposures obtained for the bright object 1867 Deiphobus. These images and all others from the same filter sequence were discarded from further analysis, and the associated metadata are not provided in the table.

\subsection{Magellan/IMACS}
\label{subsec:magellan}

\begin{deluxetable}{lcccc}[t!]
\setlength{\tabcolsep}{8pt}
\tablewidth{0pc}
\renewcommand{\arraystretch}{0.9}
\tabletypesize{\footnotesize}
\tablecaption{
    Summary of 2021 July 15--16 Magellan/IMACS Observations
    \label{tab:imacs}
}
\tablehead{
    \colhead{Object} & \colhead{UT Time\tablenotemark{\scriptsize a}} & \colhead{Airmass\tablenotemark{\scriptsize a}} & \colhead{$t_{\mathrm{exp}}$\tablenotemark{\scriptsize b}} &
    \colhead{$D_{\mathrm{aper}}$\tablenotemark{\scriptsize c}}   
}
\startdata
131451 & 03:03:26 & 1.14 & 30 & 8 \\
131460 & 02:01:39 & 1.11 & 30 & 6 \\
289501 & 09:09:33 & 1.79 & 30 & 8 \\
291276 & 04:05:09 & 1.59 & 30 & 5 \\
297019 & 02:47:45 & 1.16 & 30 & 6 \\
299491 & 07:11:53 & 1.43 & 30 & 8 \\
301010 & 03:30:16 & 1.32 & 30 & 6 \\
335567 & 23:20:48 & 1.06 & 60 & 8 \\
345407 & 04:53:31 & 1.25 & 30 & 5 \\
356934 & 02:58:58 & 1.12 & 30 & 7 \\
1997 JC11  & 01:25:13 & 1.11 & 30 & 6 \\
2003 YL133  & 04:00:15 & 1.21 & 30 & 5 \\
2005 YG204  & 00:52:13 & 1.33 & 60 & 6 \\
2007 DO47  & 00:59:56 & 1.32 & 60 & 6 \\
2007 EH99  & 02:54:36 & 1.16 & 30 & 7 \\
2007 EN217  & 03:10:23 & 1.36 & 60 & 5 \\
2007 EU219  & 05:32:58 & 1.25 & 30 & 6 \\
2008 EF7  & 07:22:34 & 1.60 & 30 & 7 \\
2008 ES68  & 07:26:54 & 1.75 & 30 & 6 \\
2008 FQ132  & 04:13:45 & 1.50 & 30 & 6 \\
2008 FV120 & 23:00:23 & 1.01 & 60 & 7 \\
2008 JR5  & 06:05:17 & 1.89 & 60 & 7 \\
2008 KE18  & 06:35:39 & 1.38 & 30 & 5 \\
2009 LJ3  & 02:08:07 & 1.05 & 60 & 5 \\
2009 MY1  & 05:49:51 & 1.58 & 30 & 5 \\
2009 SE1  & 03:24:31 & 1.20 & 30 & 5 \\
2009 SX19  & 02:14:55 & 1.37 & 60 & 6 \\
2010 HZ21  & 02:42:49 & 1.17 & 60 & 6 \\
2011 KG17  & 03:51:00 & 1.13 & 30 & 9 \\
2011 PC14  & 06:19:43 & 1.75 & 30 & 8 \\
2011 QQ64  & 03:44:15 & 1.48 & 60 & 6 \\
2011 SE216  & 07:44:48 & 1.47 & 30 & 6 \\
2012 RB27  & 04:09:33 & 1.53 & 30 & 5 \\
2012 RE39  & 05:54:42 & 1.78 & 60 & 5 \\
2012 SN6  & 02:37:58 & 1.07 & 30 & 7 \\
2012 SQ49  & 05:38:37 & 1.66 & 30 & 5 \\
2012 TD184  & 05:26:38 & 1.67 & 60 & 7 \\
2012 TD52  & 06:41:15 & 1.46 & 30 & 5 \\
2012 TE297  & 05:44:07 & 1.31 & 30 & 6 \\
2012 TM178  & 05:59:54 & 1.88 & 60 & 6 \\
2012 TP146  & 05:16:40 & 1.25 & 30 & 6 \\
2012 TR218  & 04:31:21 & 1.48 & 60 & 5 \\
2012 TX28  & 04:48:48 & 1.27 & 30 & 5 \\
2012 TY208  & 05:10:18 & 1.76 & 60 & 5 \\
2012 TZ243  & 05:03:29 & 1.27 & 60 & 9 \\
2012 UA114  & 07:16:29 & 1.40 & 30 & 7 \\
2012 US137  & 04:18:16 & 1.47 & 60 & 5 \\
\enddata
\vspace{+0cm}\textbf{Notes.}
\vspace{-0.25cm}\tablenotetext{\textrm{a}}{For each object, the start time and airmass value of the first exposure of the observing sequence(s) are provided. Objects that were not recovered due to blending and observations for which photometric zero-points could not be computed are not listed.}
\vspace{-0.15cm}\tablenotetext{\textrm{b}}{The exposure time of each image, in seconds.}
\vspace{-0.15cm}\tablenotetext{\textrm{c}}{The diameter of the circular photometric aperture (in pixels) that yields the smallest resultant color uncertainty. The pixel scale of the $2 \times 2$ binned IMACS images is $0\overset{''}{.}222$.}
\vspace{-1cm}
\end{deluxetable}

On the night of UT 2021 July 15--16, we carried out photometric observations of 69 candidate Ennomos family members with the Inamori--Magellan Areal Camera and Spectrograph (IMACS) instrument on the 6.5 m Magellan Baade Telescope at Las Campanas Observatory, Chile. The f/4 camera was used, which fully illuminates a $15\overset{'}{.}4 \times 15\overset{'}{.}4$ detector mosaic consisting of eight $2048 \times 4096$ CCDs. We selected $2 \times 2$ binning, resulting in an effective pixel scale of $0\overset{''}{.}222$, and placed the science target near the center of CCD \#8 for all exposures.

After sets of bias frames and dome flats were obtained during evening twilight, science observations proceeded in a uniform fashion: each object was visited once and imaged with a pair of exposures in the Sloan $g^{\prime}$ (4678 \AA) and $i^{\prime}$ (7655 \AA) filters. Per-exposure integration times were chosen to be either 30 or 60 s, depending on the brightness of the target. The sky remained mostly cloud-free throughout the night of observation, with intermittent patches of high-altitude wispy clouds. The measured seeing ranged from $0\overset{''}{.}5$ to $0\overset{''}{.}9$.

Several of the targets were situated in crowded fields near the galactic plane and were blended with background stars, which precluded accurate photometric extraction. Another significant fraction of objects were located at low declinations (below $-30^{\circ}$), where the availability of faint catalog stars is limited. For 15 objects, there were too few unsaturated reference stars on the CCD to allow for reliable photometric calibration (see Section~\ref{subsec:phot}). Out of the 69 objects that were imaged with Magellan/IMACS, 22 were unsuitable for color measurements. Table~\ref{tab:imacs} lists the observation details of the 47 Trojan targets for which optical colors were derived. None of these objects were previously observed with Keck/DEIMOS.

\subsection{Photometric Extraction}
\label{subsec:phot}
The data processing and photometric extraction were handled by a custom pipeline that follows standard practices for moving object imaging. For each night of observation, we generated a master bias and a set of master flats for each filter by median-averaging all individual bias and flat exposures. The pipeline then passed each bias-corrected and flat-fielded science image through SExtractor \citep{bertin1996} to generate a list of sources with their centroid positions and aperture fluxes. The astrometric solution and photometric zero-point were simultaneously calculated by automatically matching the sources to the catalog stars. The default stellar database for our reduction was the Pan-STARRS Data Release 2 catalog \citep{flewelling2020}, which contains billions of stars and their apparent magnitudes in the Pan-STARRS photometric system. To convert the Pan-STARRS magnitudes to the appropriate filter in the standard $ugriz$ or $BVRIZ$ systems, our pipeline applies the empirical transformations derived by \citet{tonry2012}. Finally, the pipeline queried the JPL Horizons database for the position of the Trojan target at the mid-exposure time, identified the corresponding source on the image, and returned the apparent magnitude and uncertainty of the object in the appropriate filter.

The Pan-STARRS catalog provides nearly complete coverage of the sky above a decl. of $-30^{\circ}$. Notably, it contains faint stars ($m > 18$), which are numerous and crucially unsaturated in our deep images. However, a significant fraction of the fields that we observed with Magellan/IMACS were located at lower declinations, which are outside the coverage area of Pan-STARRS. For these images, we relied on the Asteroid Terrestrial-impact Last Alert System (ATLAS) catalog \citep{tonry2018}. This database combines the Pan-STARRS objects with other stellar catalogs, including Tycho-2 \citep{hog2000}, the AAVSO Photometric All Sky Survey \citep{henden2016}, and SkyMapper \citep{wolf2018}, all of which have substantial coverage of the southern ecliptic sky. Conveniently, the magnitudes of all ATLAS stars are listed in the Pan-STARRS photometric system. Unfortunately, there are significantly fewer faint ($m > 18$) ATLAS stars below $-30^{\circ}$ decl. A reliable photometric zero-point calculation requires multiple unsaturated catalog stars. Given the observed spread in calculated photometric zero-points for individual sources across the Magellan/IMACS images, we stipulated a requirement of at least five reference stars per image when computing the apparent magnitude of the Trojan targets. Dozens of images from our 2021 observations did not pass this threshold and were therefore removed from consideration.

We experimented with a range of circular extraction aperture sizes to optimize the color measurements. For the Keck/DEIMOS observations, we considered aperture diameters ranging from 11 to 29 pixels in 2 pixel intervals. Meanwhile, for the Magellan/IMACS images, we selected from aperture diameters of 4--10 pixels in 1 pixel intervals. A single optimal aperture was selected for each visit (i.e., a sequence of multiband observations at a single pointing) by choosing the aperture size that minimized the resultant uncertainty in $g-i$ or $V-I$ color. The optimal aperture diameters are listed in Tables~\ref{tab:deimos} and \ref{tab:imacs} for each object.

Table~\ref{tab:colors} in the Appendix lists the measured colors for 73 candidate Ennomos family members from our photometric campaign --- 26 sets of $V-R$, $R-I$, and $V-I$ colors from the 2015/2016 Keck/DEIMOS nights; and 47 $g-i$ colors from the 2021 Magellan/IMACS observations. For objects that were observed more than once, the color measurements were combined using weighted mean averaging. In all cases, the individual values mutually agree to well within $2\sigma$. Published visible spectroscopy of dozens of Trojans has hitherto revealed uniformly featureless spectra with linear continua, including for confirmed Eurybates collisional family members \citep[e.g.,][]{dotto2006,fornasier2007,melita2008}. This allows us to straightforwardly convert the $VRI$-band measurements to $g-i$ colors. We applied the color transformations from \citet{jordi2006} to generate $g-i$ colors for the objects that were observed with Keck/DEIMOS, which are also included in Table~\ref{tab:colors}.

\section{Analysis and Discussion}
\label{sec:ana}

\subsection{Color distribution}\label{subsec:dist}

As discussed in the introduction, previous investigations into the Eurybates collisional family have demonstrated that the confirmed family members can be distinguished from the background by their closely clustered optical colors that are notably bluer than both the less-red and red subpopulations on average \citep[e.g.,][]{fornasier2007,broz2011,vinogradova2015}. Motivated by these findings, we sought to carry out an analogous analysis of the $g-i$ color distribution for the candidate Ennomos family members to filter out interloping background objects and compile a list of validated collisional family members.

We searched the Fourth Release of the SDSS Moving Object Catalog\footnote{\url{http://faculty.washington.edu/ivezic/sdssmoc/sdssmoc.html}} for objects in the candidate Ennomos family member list and found six objects. Their averaged $g-i$ colors are listed as separate entries in Table~\ref{tab:colors} and are denoted by the superscripts. Four of these objects were also observed as part of our ground-based observing campaign and the SDSS colors agree with our measurements to well within $2\sigma$ in each case. The remaining two targets --- 345407 and 348312 --- bring the total number of candidate Ennomos family members with measured colors to 75.

\begin{figure}[t]
\begin{center}
\includegraphics[width=\linewidth]{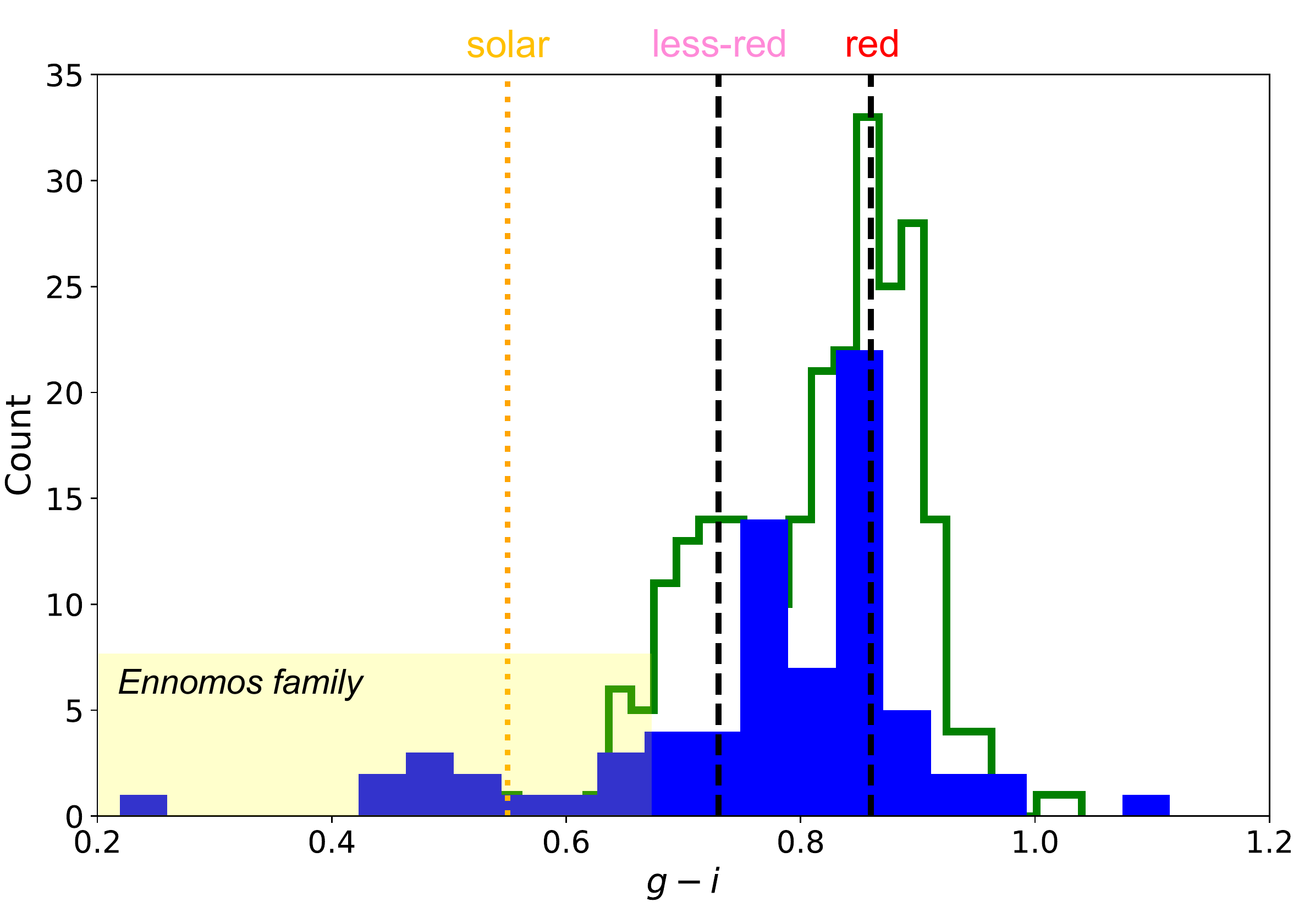}
\end{center}
\caption{Histogram of measured $g-i$ color for the purported Ennomos family members observed in our program (blue), alongside the color distribution of Trojans with $H < 12.3$ mag that were observed by the Sloan Digital Sky Survey (green). The vertical lines denote the mean $g-i$ colors for the less-red and red Trojan subpopulations, as well as the solar color. While the majority of candidate family members have colors that are indistinguishable from the background population, there is a discernible cluster of objects that are significantly bluer than the background. The yellow shaded region corresponds to $g-i < 0.68$ (i.e., the color selection criterion that we stipulate for the true Ennomos collisional family).} \label{fig:hist}
\end{figure}

\begin{figure*}[t]
\begin{center}
\includegraphics[width=\linewidth]{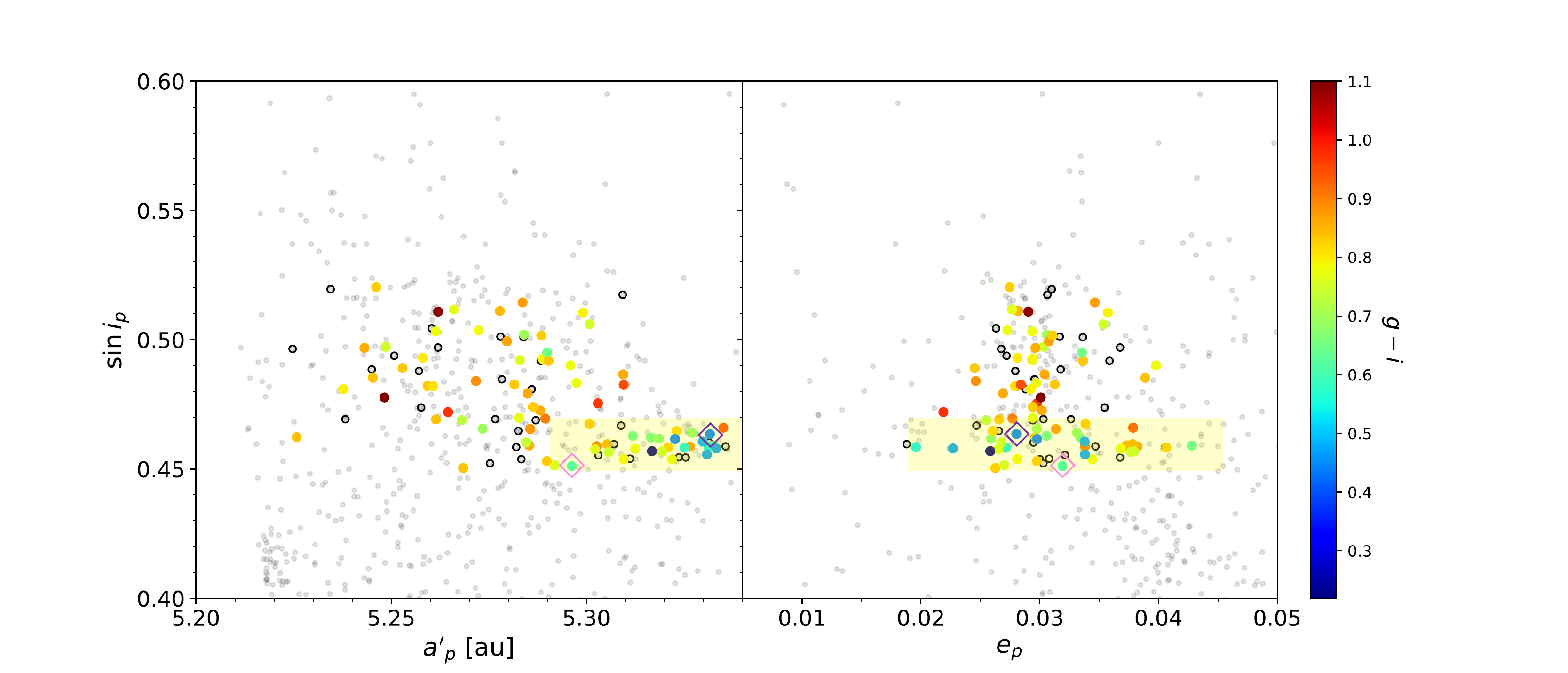}
\end{center}
\caption{Color map of the proper element distribution from Figure~\ref{fig:targets}, color-coded based on the measured $g-i$ colors. The empty white dots indicate targets that are on the \citet{rozehnal2016} candidate Ennomos family member list but were not imaged as part of our multiband photometric observations. Note the dense concentration of objects with blue--neutral colors at low proper inclination and high proper semimajor axis, which corresponds to the true extent of the Ennomos collisional family (yellow-shaded regions). The location of 4709 Ennomos and 17492 Hippasos are marked by the pink and purple diamonds, respectively.} \label{fig:colors}
\end{figure*}

In Figure~\ref{fig:hist}, we plot the histogram of measured $g-i$ colors for candidate Ennomos family members, alongside the distribution of available SDSS colors for all Trojans with $H < 12.3$ \citep{wong2014,wong2015}. The average $g-i$ colors of the less-red and red populations --- 0.73 and 0.86, respectively --- are also shown, as well as the solar color (0.55). It is apparent that most of the objects in the \citet{rozehnal2016} list of Ennomos family members have colors that are indistinguishable from the background population, with $g-i$ values spanning the full range of Trojan SDSS colors. However, a closer inspection reveals a notable cluster of objects with low $g-i$ values that are significantly bluer than the less-red subpopulation average, and in some cases are even bluer than solar.

For a more illustrative perspective, we overlay the measured $g-i$ colors on top of the corresponding proper element distribution in Figure~\ref{fig:colors}. While the majority of candidate family members have colors that are consistent with the background population, as stated previously, there is a dense kernel of objects with unusually blue colors that is confined to the high-$a^{\prime}_{p}$, low-$i_{p}$ region of the proper element space spanned by our observed targets. When comparing the location of low $g-i$ objects to the respective distributions of the candidate family members from \citet{nesvorny2015} and \citet{rozehnal2016} (Figure~\ref{fig:targets}), it is evident that the distribution of these blue--neutral objects is more consistent with the former catalog, though with a characteristic $d_{\rm cut}$ that is significantly smaller than the 100 m~s$^{-1}$ that was used in the clustering analyses (i.e., more tightly clustered). We conclude that while the majority of listed candidate Ennomos family members are evidently not part of a collisional family, there exists a subset of objects that is both (i) photometrically distinct from the background Trojan population and (ii) tightly clustered in proper element space. In the following, we use both measured colors and proper elements to designate the true extent of the Ennomos collisional family.

\subsubsection{Confirmed Ennomos family members}\label{subsec:list}

Eight objects have measured optical colors that are bluer than solar ($g-i < 0.55$) and are thereby sharply distinguishable from the background Trojan population. By examining their positions in proper element space (Figure~\ref{fig:colors}), we find that they are restricted to $0.45 < \sin{i_{p}} < 0.47$ and $a^{\prime}_{p} > 5.31$ au, while spanning a range of $e_{p}$ that is comparable to the full candidate family member lists. Given their highly distinct spectral properties and close dynamical proximity, we can confidently confirm that these eight objects are members of a collisional family: 17492 ($H = 10.14$), 98362 ($H = 12.59$), 337420 ($H = 12.79$), 187692 ($H = 12.85$), 246817 ($H = 12.85$), 2011 PC14 ($H = 13.51$), 2007 DO47 ($H = 13.63$), and 2012 TR218 ($H = 14.08$). Notably, this list includes 17492 Hippasos, which \citet{rozehnal2016} designated as the center of the dynamical cluster of objects.

We note that the measured optical color of 4709 Ennomos is slightly redder than solar ($g-i = 0.585 \pm 0.012$). By revisiting Figure~\ref{fig:colors}, we see that all but two of the 17 objects with measured $g-i$ colors less than 0.7 are grouped within the same range of proper inclinations as the eight objects listed above. For comparison, the range of $g-i$ colors for Eurybates family members in the SDSS Moving Object Catalog is 0.55--0.75. Consequently, there is a high likelihood that most of the candidate targets with $g-i < 0.7$ are also part of the Ennomos collisional family. Using the distribution of SDSS colors as the basis for our selection, we consider all objects in the range $0.45 < \sin{i_{p}} < 0.47$ that have $g-i$ colors below the fifth percentile value ($g-i = 0.68$) to be part of the Ennomos collisional family. This expanded selection criterion adds five more objects to the eight that were listed previously --- 4709 ($H = 8.76$), 301010 ($H = 13.40$), 2005 YG204 ($H = 13.55$), 2010 HZ21 ($H = 13.82$), and 2012 UA114 ($H = 14.26$) ---, which brings the total number of confirmed family members to 13.

It follows that the approximate true extent of the Ennomos collisional family in the space of proper elements and optical color is $a^{\prime}_{p} > 5.29$ au, $0.45 < \sin{i_{p}} < 0.47$, $0.019 < e_{p} < 0.045$, and $g-i < 0.68$. These ranges are delineated by the yellow shaded boxes in Figures~\ref{fig:hist} and \ref{fig:colors}.

The Hilda asteroids offer a complementary view on the interplay between collisions and surface color. The Hildas are locked in the 3:2 mean motion resonance with Jupiter and, like the Trojans, are hypothesized to be inward-scattered outer solar system bodies. In addition to displaying an analogous color bimodality to the Trojans \citep[e.g.,][]{gilhutton2008}, the Hildas are notable for containing two large collisional families --- the Hilda and Schubart families \citep{broz2008,nesvorny2015}. A previous analysis of SDSS photometry demonstrated that the color distribution of the Hilda and Schubart family members is similarly skewed to bluer colors than the background population, with the majority of family members having $g-i$ values in the same range as the Eurybates family in the Trojans \citep{wong2017hildas}. 

\citet{wong2016} explored the photometric effects of shattering collisions in the Trojan population within the framework of dynamical instability models. Primordial planetesimals in the cold outer solar system would have accreted with a generally comet-like bulk composition, including a significant amount of volatile ices (e.g., ammonia and methanol). Initial irradiation chemistry, coupled with location-dependent differential sublimation of the volatile ices, would have reddened the surfaces, with different levels of reddening based on an object's heliocentric distance. Meanwhile, the unexposed interiors would remain unaltered, with a largely uniform composition throughout the outer solar system. The result is spectrally distinct subpopulations of reddish objects, a portion of which were later scattered during the dynamical instability and implanted into the Jupiter region. This process accounts for the attested color diversity among both the Jupiter-resonant asteroid populations and the Kuiper belt. Subsequent catastrophic impacts in the Trojan population would have exposed the pristine interior, with all of the remaining volatile ices rapidly sublimating given the higher temperatures at 5.2 au. Without these volatile ices, the action of space weathering would generate a distinct surface type, which is crucially dissimilar to the weathered regolith of uncollided objects. The discovery of four major collisional families among the Jupiter-resonant asteroid populations (Eurybates, Ennomos, Hilda, Schubart) composed of relatively blue objects corroborates the predictions of this hypothesis, and more fundamentally provides strong evidence for the dynamical instability model of solar system evolution.

\subsection{Magnitude distribution}\label{subsec:mag}

\begin{figure}[t]
\begin{center}
\includegraphics[width=\linewidth]{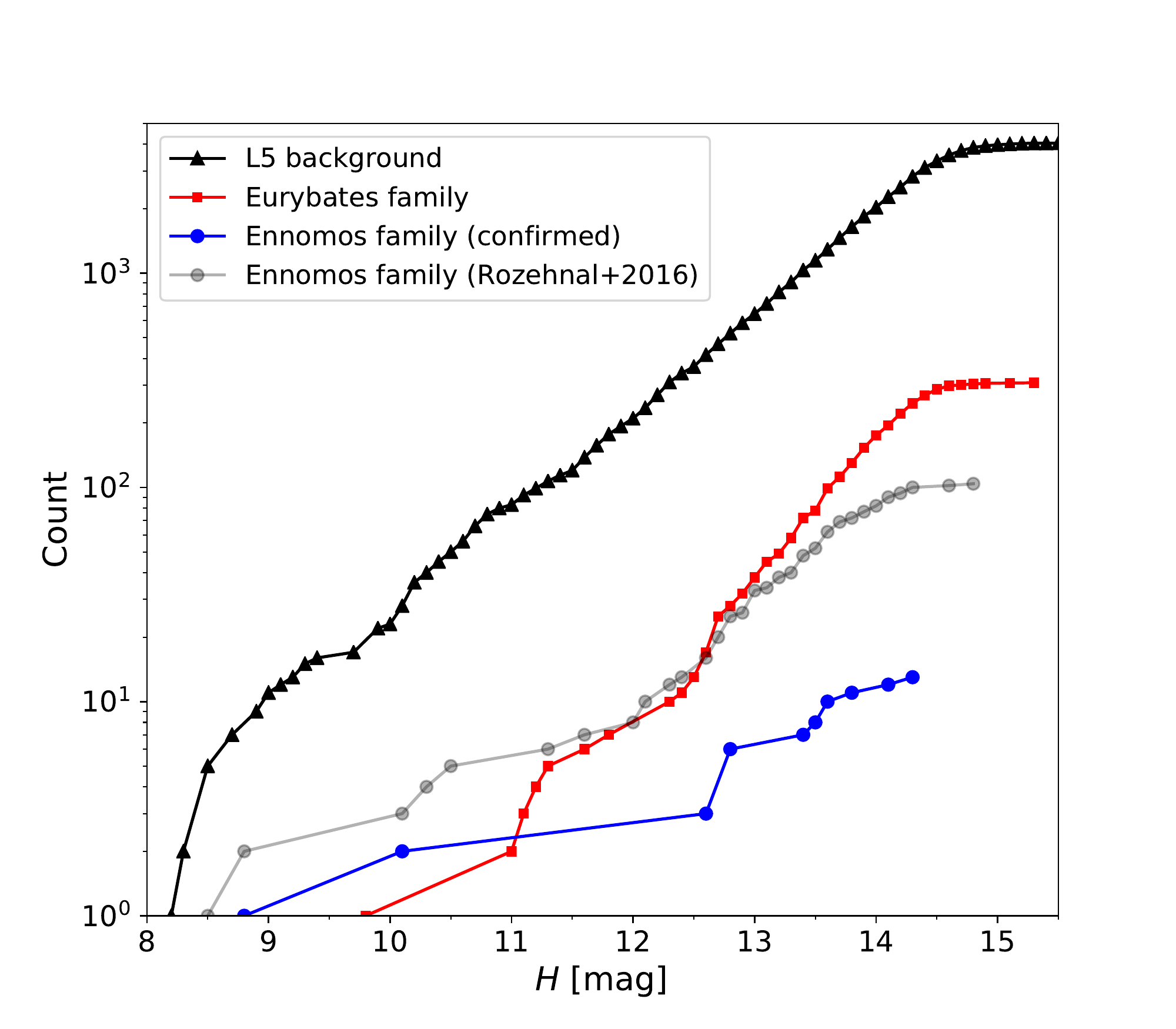}
\end{center}
\caption{Cumulative absolute magnitude distributions of the confirmed Ennomos family members (blue circles), along with the magnitude distributions of the Eurybates collisional family (red squares) and current L5 background population (black triangles). The gray circles show the magnitude distribution of the full list of 104 candidate Ennomos family members. Significant catalog incompleteness in the candidate family member datasets begins at roughly $H = 14.0$. The shape of the Ennomos family magnitude distribution is poorly constrained due to the small sample size. Meanwhile, the Eurybates family magnitude distribution is significantly steeper than the background Trojan population.} \label{fig:mag}
\end{figure}

The size distribution, or equivalently, in the case of uniform geometric albedo, the absolute magnitude distribution, is a diagnostic feature of a minor body population that reflects both the formation conditions and subsequent collisional evolution. Previous analyses have characterized the Trojan magnitude distribution down to kilometer-sized objects \citep[e.g.,][]{yoshida2008,wong2015,yoshida2017}. Generally, magnitude distributions of asteroid populations can be described with a power law $\Sigma(H) = 10^{\alpha(H-H_{0})}$, where $\Sigma(H)$ is the differential magnitude distribution, $\alpha$ is the power-law slope, and $H_{0}$ is the threshold magnitude that is used to normalize the distribution. Comparisons between the magnitude distributions of the Eurybates collisional family and the background population have revealed that the former is significantly steeper (i.e., has a higher $\alpha$ value; e.g., \citealt{broz2011}). Likewise, the characteristic $\alpha$ for the Hektor family is also steeper than that of the background population \citep[e.g.,][]{rozehnal2016}. These results indicate that the slope of the magnitude distribution may serve as a useful metric for determining whether a dynamical family has a collisional origin. 

When studying asteroid magnitude distributions, catalog incompleteness is a crucial consideration. This is especially important for our analysis of the Trojan collisional families because the proper element catalog used by \citet{rozehnal2016} was computed based on osculating elements listed in the AstOrb database \citep{bowell2002} as of 2014 July. To assess the extent of catalog incompleteness, in Figure~\ref{fig:mag} we plot the cumulative absolute magnitude distributions of the confirmed Ennomos family members and the Eurybates collisional family (both from \citealt{rozehnal2016}) alongside the current (as of this publication) L5 background population listed by the Minor Planet Center. For the sake of comparison, we also include the magnitude distribution of all 104 candidate Ennomos family members. Here, and elsewhere in this paper, we exclusively utilize absolute magnitude values from the AstOrb database for the sake of consistency. From the location of the sharp rollover in the magnitude distribution of Eurybates collisional family members, it is evident that the onset of catalog incompleteness when the \citet{rozehnal2016} candidate family member lists were derived is around $H = 14$.

Due to the small number of confirmed Ennomos family members, the fitted distribution slope is not precise enough to offer compelling independent evidence for the collisional nature of the family. Nevertheless, as part of the general characterization of the family, we carried out a fit to the magnitude distribution, following standard Monte Carlo parameter estimation techniques \citep[e.g.,][]{wong2014,wong2017hildas}. Given the limited magnitude range of the confirmed Ennomos family members and the onset of catalog incompleteness in the \citet{rozehnal2016} database, we restricted our fits to objects with absolute magnitudes between 10.5 and 14.0. For the Ennomos family, we retrieved a power-law slope of $\alpha = 0.49^{+0.20}_{-0.15}$. Meanwhile, we obtained $\alpha$ values of $0.56 \pm 0.01$ and $0.74 \pm 0.06$ for the L5 background and Eurybates family, respectively. 

The $3.0\sigma$ discrepancy in the power-law slope between the background population and Eurybates family corroborates the findings of previous studies. Meanwhile, the poorly constrained $\alpha$ value of the confirmed Ennomos family is broadly consistent with both the background and the Eurybates family. For a more rigorous assessment of the statistical distinction between the various magnitude distributions, we utilized the Kuiper variant of the two-sample Kolmogorov--Smirnov test \citep{press2007}. The probability that the Eurybates family and background population magnitude distributions are drawn from the same underlying $\Sigma(H)$ is 0.03. Meanwhile, analogous comparisons with the Ennomos family returned $p$-values above 0.2, which indicate that its magnitude distribution cannot be robustly distinguished from that of either the Eurybates collisional family or the L5 background population.

We note that there are eight candidate Ennomos family members from the \citet{rozehnal2016} list for which we do not have measured optical colors (see Section~\ref{subsec:future}). Of these, five lie within the brightness range of our magnitude distribution analysis. If we naively assume, for the sake of establishing a limiting case, that all five are in fact part of the true Ennomos collisional family, then the fitted power-law slope of the magnitude distribution rises to $\alpha = 0.58^{+0.18}_{-0.15}$. This is in closer agreement with the measured slope of the Eurybates collisional family, while remaining statistically consistent with the previously obtained value for the photometrically confirmed members alone, and formally indistinguishable from the L5 background magnitude distribution. Meanwhile, when fitting all 104 candidate Ennomos family members from \citet{rozehnal2016}, the resultant slope value is $\alpha = 0.51 \pm 0.06$. This indicates that the rejected candidate family members, which comprise the majority of the catalog and are dominated by objects at higher proper inclinations, have a magnitude distribution that is indistinguishable from the background.

There is an additional caveat that is relevant to the characterization and interpretation of Trojan magnitude distributions, particularly in the case of the Ennomos collisional family. A recent numerical analysis has revealed that the long-term dynamical stability of Trojans varies considerably across the population, with the characteristic escape timescale falling precipitously with increasing proper eccentricity, inclination, and libration amplitude \citep{holt2020}. The confirmed members of the Ennomos collisional family lie close to the edge of the stable parameter space, with relatively high proper inclinations and libration amplitudes. Consequently, a significant fraction of reference particles escaped the L5 swarm over the course of the dynamical simulations. In fact, of the six dynamical families identified by \citet{nesvorny2015}, the Ennomos family has the highest escape fraction. It is therefore likely that many fragments from the progenitor collision were ejected from the swarm and given that smaller fragments are imparted systematically higher velocities during impacts \citep[e.g.,][]{jutzi2010}, they would be preferentially lost from the L5 resonance. All else being equal, this scenario could yield a shallower size/magnitude distribution for the Ennomos family when compared to the more dynamically stable Eurybates collisional family.

The magnitude distribution of confirmed Ennomos family members is notable in having two large members --- Ennomos ($H = 8.76$) and Hippasos ($H = 10.14$) --- followed by a cluster of much smaller objects that are separated by a significant gap in size. Recent dynamical and hydrodynamical modeling of the Hobson collisional family in the Main Belt, which has a similar magnitude distribution, has demonstrated that the collisional disruption of one member of a primordial binary may explain the peculiar size distribution of family members \citep{vokrouhlicky2021}. In the case of the Ennomos family, both largest members have significantly bluer colors than the background. Consequently, in the scenario of a primordial binary, both members would have to be catastrophically disrupted. Future detailed collisional modeling, combined with a larger set of confirmed smaller fragments, may elucidate the specific formation conditions of the Ennomos family.

\subsection{Future prospects}\label{subsec:future}

We have obtained color measurements for the majority of candidate Ennomos family members from the \citet{rozehnal2016} catalog. Based on the observed distribution of true Ennomos collisional family members in proper element space (Section~\ref{subsec:dist}), any remaining candidates without published colors that are located above $\sin{i_{p}} = 0.47$ are unlikely to be associated with the family. Only eight targets lie within the region of proper element space spanned by the confirmed family members (Figure~\ref{fig:colors}): 2009 KS5 ($H = 13.10$), 2012 TQ14 ($H = 13.16$), 2012 QQ14 ($H = 13.49$), 2008 EK42 ($H = 13.71$), 2006 BY108 ($H = 13.91$), 2007 HM2 ($H = 14.15$), 2008 KY36 ($H = 14.22$), and 2001 QP331 ($H = 14.77$). Future observing campaigns to measure the colors of these targets will solidify the shape of the Ennomos family magnitude distribution through $H= 14$.

Nevertheless, even if all or most of these uncharacterized objects are confirmed to be collisional fragments of the Ennomos impact, the total number of family members would still be insufficient to allow for precise characterization of the magnitude distribution. The photometric study of fainter objects is necessary to refine our understanding of the Ennomos collisional family, as well as the broader Trojan population. In the years that have transpired since the creation of the extant Trojan dynamical family catalogs, thousands of new objects have been discovered, most with $H > 14$. Indeed, the most recent catalog of Trojan proper elements \citep{holt2020b} has filled in numerous smaller objects in the general region of the Ennomos family, as illustrated in Figure~\ref{fig:colors}. The upcoming decade-long Legacy Survey of Space and Time (LSST) at the Rubin Observatory \citep{izevic2019}, with a $5\sigma$ single-exposure limiting magnitude of $\sim$24, will potentially uncover hundreds of thousands of faint Trojans down to $H \sim $~17--18. By reexamining the dynamical landscape of the Trojan swarms, a drastically more complete picture of the number and extent of collisional families will emerge. LSST will provide multiband visible-wavelength photometry of all imaged sources, which will enable simultaneous characterization of the faint Trojan color distribution and effective interloper rejection among proposed collisional families through the use of similar techniques to those outlined in this paper.

While the colors of Trojan asteroids are useful for classifying objects into the less-red and red subpopulations, and for probing collisional family membership, they do not directly provide any definitive information about the physical and chemical properties of the surfaces. Compositional modeling of high-quality reflectance spectra is indispensable in this regard. While previously published spectroscopic datasets spanning the near-ultraviolet, optical, and near-infrared have largely revealed featureless spectra, next-generation instruments are poised to contribute incisive fresh insights into our foundational understanding of Trojan formation and evolution. 

After initiating science observations earlier this year, the James Webb Space Telescope (JWST) will become the workhorse facility in these endeavors. The unique strengths of the JWST instrument suite in the realm of small body astronomy are particularly salient in the near-infrared, which is a challenging wavelength range for ground-based observations that hosts key spectral signatures of most major regolith elements (e.g., water ice, volatile ices, silicates, salts, and organics). By leveraging the continuous wavelength range of NIRSpec from 0.8 through 5 microns, an approved Cycle 1 proposal (PI: M. Brown) will obtain high signal-to-noise ratio reflectance spectra of all five Trojan flyby targets of the Lucy mission at unprecedented spectral resolution. An analogous future proposal targeting some of the brightest confirmed Ennomos family members has the potential to greatly broaden the impact of JWST in enriching our knowledge of Trojan asteroids. Our analysis has uncovered a surprising cluster of bluer-than-solar objects, including 337420 ($g-i = 0.219 \pm 0.014$), which is one of the bluest objects known in the middle and outer solar system. The distinct colors of the Ennomos family suggest a unique surface composition. With both reflectance and emission spectra of Eurybates soon becoming available through the aforementioned approved Cycle 1 program, a detailed spectral comparison of two different collisional families would offer novel constraints on the diversity of interior compositions within the Trojan population.

Looking further into the future, the 2027--2033 Lucy flybys will establish the definitive paradigm for Trojan chemical composition, surface properties, and dynamical history \citep{levison2021}. In particular, the exquisite, spatially resolved observations of Eurybates will serve as the template for our understanding of collisional families throughout the Trojan population.

\section{Summary}
\label{sec:con}

We conducted a photometric survey of candidate Ennomos family members using the Keck/DEIMOS and Magellan/IMACS instruments in 2015/2016 and 2021. The primary findings of our analysis are summarized below:
\begin{enumerate}
 \item Out of the 104 cataloged Ennomos family members that were compiled by \citet{rozehnal2016}, we obtained $g-i$ colors of 75 objects (Table~\ref{tab:colors}). Most of these targets have colors that are consistent with the background population and are therefore not part of the true collisional family. Meanwhile, we identified a small group of objects that are significantly bluer than the background ($g-i < 0.68$) and tightly clustered in the space of proper elements. These 13 objects, listed in Section~\ref{subsec:list}, are the confirmed Ennomos collisional family members.
 \item Due to the small number of confirmed members, the magnitude distribution of the Ennomos family is poorly characterized, with a power-law slope of $0.49^{+0.20}_{-0.15}$. The distribution is statistically indistinguishable from both the Eurybates family and the L5 background population. The confirmation of additional smaller family members is necessary to better constrain the shape of the magnitude distribution.
 \item The Ennomos family joins the Eurybates family, as well as the Hilda and Schubart collisional families within the Hilda asteroids, in having distinctively bluer-than-background surface colors. This emergent trend provides evidence against in situ formation of the Jupiter-resonant asteroid populations and aligns with the predictions of dynamical instability models of solar system evolution.
\end{enumerate}

\quad
\quad

This paper includes data gathered with the 6.5 m Magellan Telescopes located at Las Campanas Observatory, Chile. Some of the data presented herein were obtained at the W.M.~Keck Observatory, which is operated as a scientific partnership among the California Institute of Technology, the University of California and the National Aeronautics and Space Administration. The Observatory was made possible by the generous financial support of the W.M.~Keck Foundation. The authors wish to recognize and acknowledge the very significant cultural role and reverence that the summit of Maunakea has always had within the indigenous Hawaiian community.  We are most fortunate to have the opportunity to conduct observations from this mountain. I.W. is supported by an appointment to the NASA Postdoctoral Program at the NASA Goddard Space Flight Center, administered by Oak Ridge Associated Universities under contract with NASA. We thank Miroslav Bro{\v z} for detailed comments during the revision process that greatly improved the manuscript.

\facilities{Keck/DEIMOS, Magellan/IMACS.}
\software{SExtractor \citep{bertin1996}.}

\appendix

\section{Optical Colors of Candidate Ennomos Family Members}
Table~\ref{tab:colors} lists the optical colors of 75 candidate Ennomos family members. Of these, 73 were observed as part of our 2015/2016 and 2021 observations. We also include the $g-i$ colors of six objects that are contained in the SDSS Moving Object Catalog. For the four targets with both SDSS and Keck/DEIMOS or Magellan/IMACS photometry, we list the respective color measurements separately. The more precise values from our ground-based campaign were used in the color distribution analysis (Section~\ref{subsec:dist}). The measured $VRI$-band colors of the Keck/DEIMOS targets were converted to $g-i$ colors using the color transformations of \citet{jordi2006}.

\startlongtable
\begin{deluxetable*}{lcccc}
\setlength{\tabcolsep}{12pt}
\tablewidth{0pc}
\renewcommand{\arraystretch}{0.9}
\tabletypesize{\small}
\tablecaption{
    Measured Colors of Candidate Ennomos Family Members
    \label{tab:colors}
}
\tablehead{
    \colhead{Object} & \colhead{$V-R$} & \colhead{$R-I$} & \colhead{$V-I$} &
    \colhead{$g-i$}   
}
\startdata
1867\tablenotemark{\scriptsize a} & $\dots$ & $\dots$ & $\dots$ & $0.83 \pm 0.02$ \\
4709 & $0.353 \pm 0.005$ & $0.352 \pm 0.004$ & $0.705 \pm 0.005$ & $0.585 \pm 0.012$ \\
17492 & $0.256 \pm 0.021$ & $0.365 \pm 0.021$ & $0.622 \pm 0.011$ & $0.444 \pm 0.040$ \\
32461 & $0.409 \pm 0.008$ & $0.452 \pm 0.007$ & $0.860 \pm 0.008$ & $0.773 \pm 0.017$ \\
36624 & $0.393 \pm 0.004$ & $0.446 \pm 0.003$ & $0.838 \pm 0.004$ & $0.742 \pm 0.011$ \\
48373 & $0.481 \pm 0.011$ & $0.510 \pm 0.009$ & $0.991 \pm 0.011$ & $0.946 \pm 0.022$ \\
55419 & $0.400 \pm 0.005$ & $0.431 \pm 0.005$ & $0.832 \pm 0.005$ & $0.738 \pm 0.012$ \\
69437 & $0.466 \pm 0.010$ & $0.536 \pm 0.013$ & $0.991 \pm 0.013$ & $0.947 \pm 0.023$ \\
76867 & $0.425 \pm 0.004$ & $0.489 \pm 0.003$ & $0.913 \pm 0.004$ & $0.836 \pm 0.011$ \\
77894 & $0.436 \pm 0.005$ & $0.474 \pm 0.005$ & $0.909 \pm 0.005$ & $0.837 \pm 0.013$ \\
98362 & $0.324 \pm 0.009$ & $0.284 \pm 0.012$ & $0.607 \pm 0.013$ & $0.472 \pm 0.021$ \\
98362\tablenotemark{\scriptsize a} & $\dots$ & $\dots$ & $\dots$ & $0.54 \pm 0.07$ \\
122592 & $0.426 \pm 0.006$ & $0.459 \pm 0.005$ & $0.883 \pm 0.006$ & $0.808 \pm 0.014$ \\
122592\tablenotemark{\scriptsize a} & $\dots$ & $\dots$ & $\dots$ & $0.87 \pm 0.04$ \\
131451 & $\dots$ & $\dots$ & $\dots$ & $0.833 \pm 0.016$ \\
131460 & $\dots$ & $\dots$ & $\dots$ & $0.967 \pm 0.022$ \\
154417 & $0.379 \pm 0.006$ & $0.372 \pm 0.006$ & $0.750 \pm 0.006$ & $0.646 \pm 0.014$ \\
187692 & $0.323 \pm 0.007$ & $0.294 \pm 0.007$ & $0.617 \pm 0.007$ & $0.480 \pm 0.015$ \\
188976 & $0.469 \pm 0.010$ & $0.445 \pm 0.014$ & $0.914 \pm 0.016$ & $0.862 \pm 0.023$ \\
215319 & $0.424 \pm 0.006$ & $0.490 \pm 0.006$ & $0.912 \pm 0.007$ & $0.835 \pm 0.015$ \\
215542 & $0.392 \pm 0.018$ & $0.490 \pm 0.016$ & $0.882 \pm 0.020$ & $0.784 \pm 0.034$ \\
246817 & $0.306 \pm 0.008$ & $0.286 \pm 0.008$ & $0.594 \pm 0.010$ & $0.446 \pm 0.017$ \\
247967 & $0.391 \pm 0.013$ & $0.482 \pm 0.015$ & $0.873 \pm 0.015$ & $0.774 \pm 0.027$ \\
247967\tablenotemark{\scriptsize a} & $\dots$ & $\dots$ & $\dots$ & $0.81 \pm 0.09$ \\
284226 & $0.503 \pm 0.013$ & $0.503 \pm 0.012$ & $1.004 \pm 0.013$ & $0.973 \pm 0.025$ \\
287454 & $0.387 \pm 0.012$ & $0.488 \pm 0.010$ & $0.871 \pm 0.013$ & $0.775 \pm 0.023$ \\
289501 & $\dots$ & $\dots$ & $\dots$ & $0.778 \pm 0.023$ \\
291276 & $\dots$ & $\dots$ & $\dots$ & $0.782 \pm 0.018$ \\
293486 & $0.398 \pm 0.007$ & $0.455 \pm 0.007$ & $0.853 \pm 0.008$ & $0.759 \pm 0.016$ \\
297019 & $\dots$ & $\dots$ & $\dots$ & $0.870 \pm 0.024$ \\
299491 & $\dots$ & $\dots$ & $\dots$ & $0.836 \pm 0.018$ \\
301010 & $\dots$ & $\dots$ & $\dots$ & $0.671 \pm 0.020$ \\
321706 & $0.445 \pm 0.015$ & $0.500 \pm 0.012$ & $0.944 \pm 0.015$ & $0.879 \pm 0.028$ \\
335567 & $\dots$ & $\dots$ & $\dots$ & $0.904 \pm 0.025$ \\
337420 & $0.217 \pm 0.006$ & $0.201 \pm 0.006$ & $0.417 \pm 0.007$ & $0.219 \pm 0.014$ \\
345407 & $\dots$ & $\dots$ & $\dots$ & $0.839 \pm 0.021$ \\
345407\tablenotemark{\scriptsize a} & $\dots$ & $\dots$ & $\dots$ & $0.83 \pm 0.04$ \\
348312\tablenotemark{\scriptsize a} & $\dots$ & $\dots$ & $\dots$ & $0.79 \pm 0.05$ \\
356934 & $\dots$ & $\dots$ & $\dots$ & $1.278 \pm 0.024$ \\
1997 JC11 & $\dots$ & $\dots$ & $\dots$ & $0.795 \pm 0.023$ \\
2003 YL133 & $\dots$ & $\dots$ & $\dots$ & $0.832 \pm 0.011$ \\
2005 YG204 & $\dots$ & $\dots$ & $\dots$ & $0.645 \pm 0.026$ \\
2005 YG204\tablenotemark{\scriptsize a} & $\dots$ & $\dots$ & $\dots$ & $0.53 \pm 0.11$ \\
2006 BK240 & $0.437 \pm 0.012$ & $0.483 \pm 0.009$ & $0.920 \pm 0.012$ & $0.849 \pm 0.023$ \\
2007 DO47 & $\dots$ & $\dots$ & $\dots$ & $0.540 \pm 0.033$ \\
2007 EH99 & $\dots$ & $\dots$ & $\dots$ & $0.862 \pm 0.037$ \\
2007 EN217 & $\dots$ & $\dots$ & $\dots$ & $0.836 \pm 0.031$ \\
2007 EU219 & $\dots$ & $\dots$ & $\dots$ & $0.680 \pm 0.023$ \\
2008 EF7 & $\dots$ & $\dots$ & $\dots$ & $0.751 \pm 0.020$ \\
2008 ES68 & $\dots$ & $\dots$ & $\dots$ & $0.833 \pm 0.076$ \\
2008 FD133 & $0.418 \pm 0.011$ & $0.467 \pm 0.010$ & $0.886 \pm 0.011$ & $0.803 \pm 0.022$ \\
2008 FQ132 & $\dots$ & $\dots$ & $\dots$ & $0.859 \pm 0.044$ \\
2008 FV120 & $\dots$ & $\dots$ & $\dots$ & $0.877 \pm 0.044$ \\
2008 JR5 & $\dots$ & $\dots$ & $\dots$ & $0.831 \pm 0.037$ \\
2008 KE18 & $\dots$ & $\dots$ & $\dots$ & $0.694 \pm 0.124$ \\
2009 LJ3 & $\dots$ & $\dots$ & $\dots$ & $0.762 \pm 0.021$ \\
2009 MY1 & $\dots$ & $\dots$ & $\dots$ & $0.740 \pm 0.022$ \\
2009 SE1 & $\dots$ & $\dots$ & $\dots$ & $0.885 \pm 0.018$ \\
2009 SX19 & $\dots$ & $\dots$ & $\dots$ & $0.832 \pm 0.026$ \\
2010 HZ21 & $\dots$ & $\dots$ & $\dots$ & $0.616 \pm 0.029$ \\
2011 KG17 & $\dots$ & $\dots$ & $\dots$ & $1.089 \pm 0.023$ \\
2011 PC14 & $\dots$ & $\dots$ & $\dots$ & $0.472 \pm 0.024$ \\
2011 QQ64 & $\dots$ & $\dots$ & $\dots$ & $0.905 \pm 0.028$ \\
2011 SE216 & $\dots$ & $\dots$ & $\dots$ & $0.834 \pm 0.036$ \\
2012 QK22 & $0.439 \pm 0.012$ & $0.459 \pm 0.011$ & $0.899 \pm 0.013$ & $0.828 \pm 0.024$ \\
2012 RB27 & $\dots$ & $\dots$ & $\dots$ & $0.788 \pm 0.023$ \\
2012 RE39 & $\dots$ & $\dots$ & $\dots$ & $0.823 \pm 0.043$ \\
2012 SN6 & $\dots$ & $\dots$ & $\dots$ & $0.858 \pm 0.022$ \\
2012 SQ49 & $\dots$ & $\dots$ & $\dots$ & $0.867 \pm 0.028$ \\
2012 TD184 & $\dots$ & $\dots$ & $\dots$ & $0.699 \pm 0.049$ \\
2012 TD52 & $\dots$ & $\dots$ & $\dots$ & $0.795 \pm 0.025$ \\
2012 TE297 & $\dots$ & $\dots$ & $\dots$ & $0.750 \pm 0.021$ \\
2012 TM178 & $\dots$ & $\dots$ & $\dots$ & $0.778 \pm 0.046$ \\
2012 TP146 & $\dots$ & $\dots$ & $\dots$ & $0.860 \pm 0.021$ \\
2012 TR218 & $\dots$ & $\dots$ & $\dots$ & $0.538 \pm 0.034$ \\
2012 TX28 & $\dots$ & $\dots$ & $\dots$ & $0.856 \pm 0.037$ \\
2012 TY208 & $\dots$ & $\dots$ & $\dots$ & $0.780 \pm 0.036$ \\
2012 TZ243 & $\dots$ & $\dots$ & $\dots$ & $0.725 \pm 0.068$ \\
2012 UA114 & $\dots$ & $\dots$ & $\dots$ & $0.637 \pm 0.047$ \\
2012 US137 & $\dots$ & $\dots$ & $\dots$ & $0.749 \pm 0.028$ \\
\enddata
\vspace{+0.1cm}\textbf{Note.}
\vspace{-0.15cm}\tablenotetext{\textrm{a}}{These colors are derived from SDSS photometry. For objects with both SDSS and Keck/DEIMOS or Magellan/IMACS colors, the latter are used in our color distribution analysis due to their higher precision.}
\end{deluxetable*}

{}
\end{document}